\documentclass[a4paper]{jpconf}
\usepackage{graphicx,psfrag}
\usepackage{epsfig}
\usepackage{amsmath}
 \usepackage{amscd}
 \usepackage{amssymb}
\usepackage{epsf,float}
\usepackage{bm}%

\newcommand{\pg}{\psfrag}

\begin{document}
\title{de Broglie Deterministic Dice \\ and  emerging Relativistic Quantum Mechanics}

\author{Donatello Dolce}

\address{  
 The University of Melbourne \& CoEPP, School of Physics, Parkville VIC 3010, 
Australia.}

\ead{ddolce@unimelb.edu.au}

\begin{abstract}
Generalizing  de Broglie's hypothesis, we show that the basic quantum  behavior of ordinary field theory can be retrieved in a semi-classical and geometrical way from the assumption of intrinsic periodicity of elementary systems. The geometrodynamical description of interactions that arises in from this approach provides an intuitive interpretation of  Maldacena's conjecture and it turns out to be of the same type of the one prescribed by general relativity.    
\end{abstract}

\section{Introduction}

Similarly to a particle in a box or to a vibrating string, relativistic fields can be quantized by imposing their characteristic de Broglie periodicities as constraints \cite{Dolce:2009ce,Dolce:tune,Dolce:AdSCFT,Dolce:2009cev4,Dolce:QTRF5,Dolce:2010ij,Dolce:FQXi}. 
In this way elementary systems can  be thought of as  "de Broglie internal clocks", that is, as relativistic fields with intrinsic dynamical de Broglie time periodicities  $T_{t}$. 
Even for a light particle such as the electron, this intrinsic time periodicity (also know as \emph{zitterbewegung}) is extremely fast ($T_{t} \lesssim 10^{-21} s$), many orders of magnitude away from the present experimental time resolution.  
As in \emph{dice} rolling too fast with respect to a given time resolution, these de Broglie internal clocks can only be  described statistically. 
It can be shown that the effective statistical description  emerging from such periodic dynamics matches (without fine tunings) ordinary relativistic Quantum Mechanics (QM) \cite{Dolce:2009ce}. 
The idea is similar to  the "stroboscopic quantization" \cite{Elze:2003tb} or to the 't Hooft determinism \cite{'tHooft:2001ar}.   
At the same time the underling classical-relativistic physics seems to solve fundamental conceptual difficulties of the canonical quantum theory. 
In particular they do not involve any local-hidden-variable so that we can actually speak about deterministic quantization. 
Our assumption of dynamical periodic fields can be regarded as a combination of  Newton's  law of inertia and the de Broglie-Planck hypothesis of periodic matter waves: \textit{elementary isolated systems must be supposed to have  persistent periodicities as long as they do not interact}. 
To the energy of an elementary system $\bar E = \hbar \bar \omega$ we associate an intrinsic time periodicity is $T_{t}= 2\pi/\bar \omega$. 
Furthermore  $T_{t}$ must be considered to be dynamical, since it is related to the energy through the de Broglie like relation  $T_{t}=h/\bar E$. 
This dynamical behavior of the periodicities guarantees full consistence with relativistic causality and time ordering. 
Interactions can be introduced in the theory  as variations of the de Broglie periodicities of the fields. 
They can be equivalently formulated in terms of geometric deformations of the compact space-time dimensions of fields with Periodic Boundary Conditions
(PBCs), in a way that mimics the same geometrodynamics of General Relativity (GR).  
The good behaviors of the theory arise from the fact that it adds a cyclic property to the usual notion of relativistic time. 
After all time can only be defined by counting the number of cycles of isolated phenomena \textit{supposed} to be periodic\footnote{``\textit{By a clock we understand anything characterized by a phenomenon passing periodically through
identical phases so that we must assume, by the principle of sufficient reason, that all that
happens in a given period is identical with all that happens in an arbitrary period.}'' ~~~~~~~~~~~~~~~~~~~~~~~~~~~~~~~~~~~~~~~~~~~~~~~~~~~~~~~~~~~~~~~ A. Einstein \cite{Einstein:1910}} such as the Cs atomic clock ($T_{Cs} \sim 10^{-10} s$).  
This approach turns out to be of particular interest in addressing the problem of the arrow of time.

\section{Cyclic fields}

	The relativistic generalization of  Newton's law of inertia states that every isolated elementary system has  persistent four-momentum $\bar p_{\mu}=\{\bar E/c,\mathbf{\bar p}\}$. On the other hand,  the de Broglie-Planck formulation of QM prescribes that the four-momentum must be associated with the four-angular-frequency $\bar \omega_{\mu} = \bar p_{\mu} c / \hbar$ of a corresponding field. Here we will assume that every elementary system is described in terms of intrinsically periodic fields whose periodicities are the usual de Broglie-Planck periodicities $T^{\mu}=\{T_{t},\vec \lambda_{x} / c \}$.  As 
 Newton's law of inertia doesn't imply that every point particle moves on a straight line, our assumption of intrinsic periodicities does not mean that
the physical world should appear to be periodic. 
In fact, the four-periodicity $T^{\mu}$ is fixed dynamically by the four-momentum through the de Broglie-Planck relation  
\begin{equation} 
c \bar p_{\mu} T^{\mu} = {h} \,. \label{deBroglie:Planck:rel}
\end{equation}
 Thus the variation of four-momentum that occurs during interactions corresponds to a retarded variation of the intrinsic periodicities of the fields. This will guarantee time ordering and relativistic causality. 

Considering for the sake of simplicity only the case of periodicity $T_{t}(\mathbf{p})$ along  the time coordinate, the cyclic field  turns out to be  decomposed as a tower of frequency eigenmodes
 \begin{equation}
\Phi(\mathbf{x},t)=\sum_{n} A_{n} \phi_{n}(\mathbf{x}) u_{n}(t)~,~~~~~~~~ \text{where}~~~~ ~~u_{n}(t)=e^{-i \omega_{n}(\mathbf{\bar p}) t}\,. \label{field:exp:modes}
 \end{equation}
Its quantized angular frequency spectrum is
 \begin{equation}
\omega_{n}(\mathbf{p}) = n \bar \omega(\mathbf{\bar p})= n \frac{2 \pi}{T_{t} (\mathbf{\bar p})}\,.
 \end{equation}
 The eigenmodes correspond to the harmonic modes of a string vibrating in a compact dimension with compactification length $T_{t}(\mathbf{p})$ and PBCs, \textit{i.e.} in a cyclic dimension. 
 Such a system can be imagined as a so called ``de Broglie internal clock'', that is, a periodic phenomenon whose time periodicity is fixed dynamically by the inverse of its energy
  $
 T_{t}(\mathbf{p}) = {h}/{\bar E(\mathbf{p})}
  $. 
Thus, according to the de Broglie-Planck relation eq.(\ref{deBroglie:Planck:rel}), we have a quantized energy spectrum 
  \begin{equation}
 E_{n}(\mathbf{p}) = n \bar E(\mathbf{p}) = n \frac{h}{T_{t}(\mathbf{p})} \,.
  \end{equation}
 The dependence on the spatial momentum $\mathbf{\bar p}$ means that the time periodicity depends on the reference frame.  Indeed $T_{t}(\mathbf{p})$ is a time interval, it transforms dynamically according to Lorentz covariance.  
 
 In a relativistic framework it is easy to figure out that time periodicity induces periodicities on the spatial dimensions as well. They must necessarily be  considered  to have a consistent Lorentz covariant theory. Since in this case the whole physical information of the system is contained in a single four-period $T^{\mu}$, our intrinsically four-periodic free field can be described by the following action in compact four dimensions with PBCs 
  \begin{equation}
 \mathcal S_{\lambda_{s}} = \oint_{0}^{T^{\mu}} d^{4} x \mathcal L_{\lambda_{s}} (\partial_{\mu}\Phi, \Phi)\,.\label{free:act}
  \end{equation}
 It is important to note that PBCs (represent by the circle in the volume integral symbol) 
 minimize the action at the boundaries, in particular of the compact time dimension. Therefore they have the same formal validity as the usual (Synchronous) BCs assumed in ordinary field theory (fixed variations at the time boundaries, \textit{i.e.} $\delta \Phi |_{\Sigma} = 0$). This is an essential feature  because it guarantees that all the symmetries of the relativistic theory are preserved (as in usual field theory), even though we are assuming PBCs. In particular it guarantees that the theory is Lorentz covarince (the theory is based upon relativistic waves with retarded propagators \cite{Dolce:2009ce}). 
 For instance we can assume a generic Lorentz transformation  
  \begin{equation}
 d x^{\mu} \rightarrow   d x'^{\mu} = \Lambda^{\mu}_{\nu} ~ d x^{\nu} ~,~~~~~~~~~~ \bar p_{\mu} \rightarrow  \bar p'_{\mu} = \Lambda_{\mu}^{\nu}  ~ \bar p_{\nu} \,.\label{space:mom:Lorentz:tranf}
 \end{equation}
By definition, the four-periodicity $T^{\mu}$ is  such that the phase of the field is invariant under four-periodic translations
 $
\exp{[-i x^{\mu} \bar p_{\mu}/\hbar]} = \exp{[- i (x^{\mu} + c T^{\mu}) \bar p_{\mu}/\hbar]}\,.  
 $
 In this way we see that the four-periodicity is actually a contravariant four-vector. It 
 transforms under Lorentz transformations in the following way
  \begin{equation}
 T^{\mu} \rightarrow  T'^{\mu} = \Lambda^{\mu}_{\nu} ~  T^{\nu}\label{period:Lorentz:tranf}
 \end{equation}
so that the phase of the field is a scalar quantity under Lorentz transformations - de Broglie phase harmony. 
This can also be  inferred by noticing that after the transformation of variables eq.(\ref{space:mom:Lorentz:tranf}), the free action eq.(\ref{free:act})  turns out to be
 \begin{equation}
 \mathcal S_{\lambda_{s}} = \oint_{0}^{T'^{\mu}} d^{4} x' \mathcal L_{\lambda_{s}} (\partial'_{\mu}\Phi, \Phi)\,.\label{Lorentz:trans:action}
  \end{equation}
Therefore, in the new reference system, the new four-periodicity $T'^{\mu}$ of the field is actually given by eq.(\ref{period:Lorentz:tranf}), that is, eq.(\ref{Lorentz:trans:action}) describes a system with  four-momentum $\bar p'_{\mu}$  given by eq.(\ref{space:mom:Lorentz:tranf}).  

 The proper-time  periodicity of  massive particles, also known as the ``de Broglie periodic phenomenon'', is represented by $ T_{\tau} = T_{t}(0)$.  Since $T_{\mu}$ is a tangent  vector, we can introduced the mnemonic notation $c \bar p_{\mu}  = h / T^{\mu}$, so that it is easy to figure out that it transforms in the relativistic way according to the following constraint
  \begin{equation}
\frac{1}{T_{\tau}^{2}} =  \frac{1}{T_{t}^{2}(\mathbf{p})} - \frac{c^{2}}{\vec \lambda_{x}^{2}(\mathbf{p})}=\frac{1}{T_{\mu}}\frac{1}{T^{\mu}}\,.\label{four-period:contraint}
  \end{equation}
It can be derived for instance by dividing  the relativistic dispersion relation 
 $ 
\bar M^{2} c^{4}=\bar E^{2}(\mathbf{\bar p}) - \mathbf{\bar p}^{2} c^{2} = \bar p^{\mu} \bar p_{\mu} c^{2}
 $ 
 by the Planck constant, \emph{i.e.} by using eq.(\ref{deBroglie:Planck:rel}).
 
\begin{figure}[h]
\pg{E}{$ E_{n}(\mathbf{\bar{p}})$}
\pg{R}{$ T_{t}(\mathbf{\bar{p}})$}
\pg{0.5}{}
\pg{1}{}
\pg{m}{\!\!\!\!\!\!\!\!${\bar M} c^{2}$}
\pg{k}{\!\!\!\!\!\!\!\!\!\!\!\!$n \bar M c^{2}$}
\pg{l}{$T_{\tau}= T_{t} (0) =\frac{\lambda_{s}}{ c} =\frac{h}{\bar M c^{2}}$}
\pg{w}{$\hbar  \bar \omega(\mathbf{\bar p})$}
\pg{p}{$\mathbf {\bar p}$}
\pg{n}{$n \hbar  \bar \omega(\mathbf{\bar p})$}
\begin{center}
\begin{minipage}{14pc}
$a)$ \includegraphics[width=14pc]{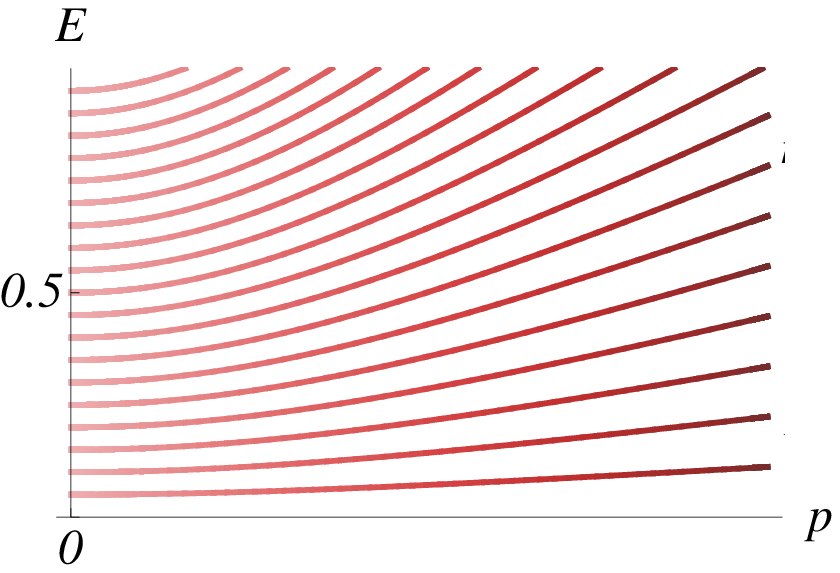}
\end{minipage}\hspace{6pc}%
\begin{minipage}{14pc}
$b)$ \includegraphics[width=14pc]{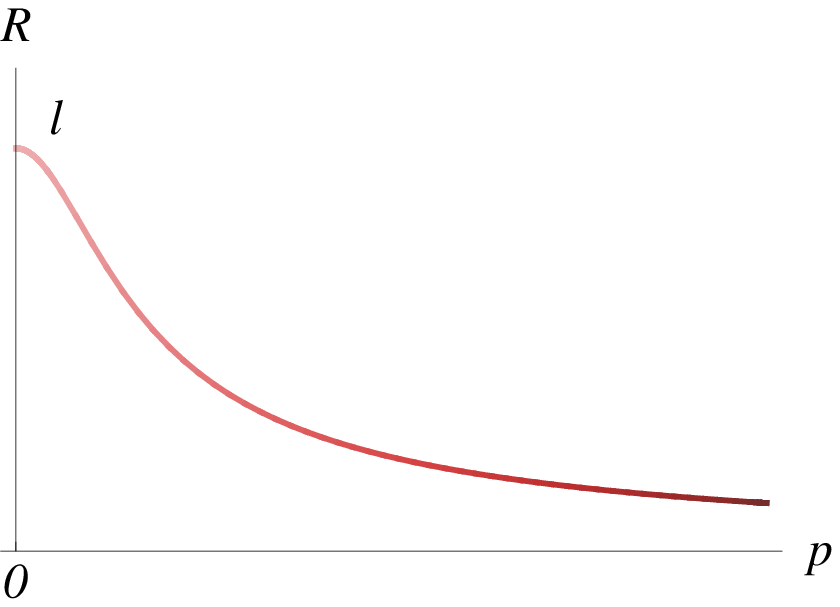}
\end{minipage} 
\begin{minipage}{14pc}
$c)$ \includegraphics[width=14pc]{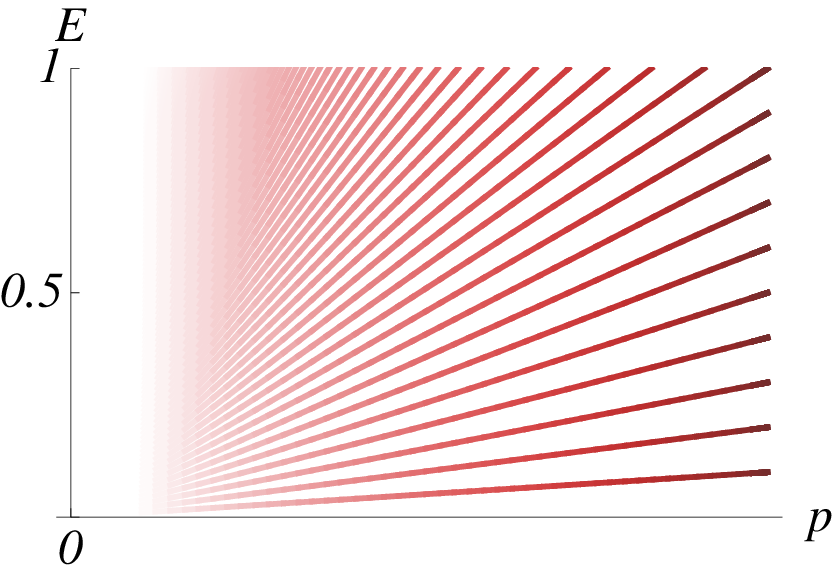}
\end{minipage}\hspace{6pc}%
\begin{minipage}{14pc}
$d)$ \includegraphics[width=14pc]{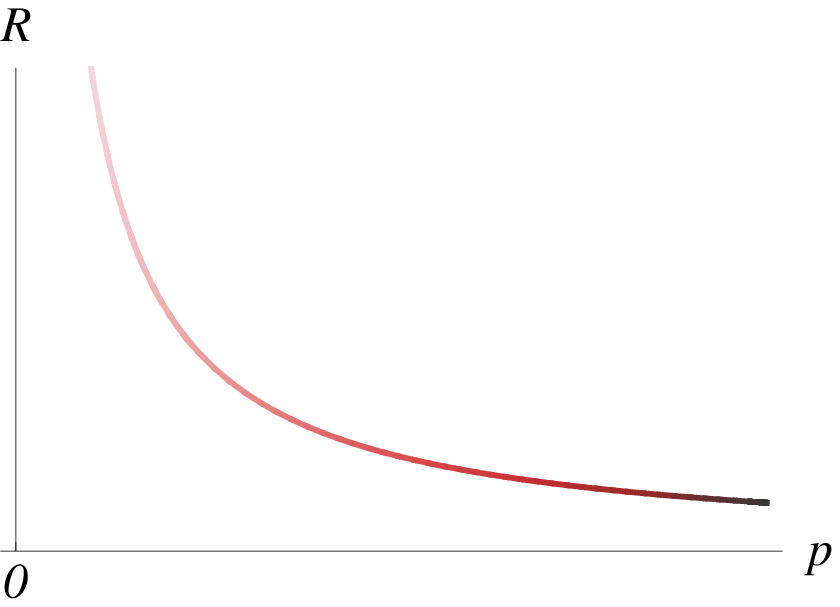}
\end{minipage}
\caption{\label{fig:disp:rel} Dispersion relation $a)$  - $c)$ -  of the quantized energy spectrum $E_{n}(\mathbf p)$  and $b)$ - $d)$ - the corresponding time periodicity $T_{t}(\mathbf p)$   of a massive - massless - cyclic field.} 
\end{center}
\end{figure}

  If we give energy to a field through interaction or we move away from the rest frame we have a relativistic deformation  of its intrinsic time periodicity, as shown in fig.(\ref{fig:disp:rel}.b) and eq.(\ref{period:Lorentz:tranf}). Therefore every level of the quantized energy spectrum transforms according to a relativistic dispersion relation. The  resulting quantized energy spectrum of the field is
 \begin{equation}
E_{n}(\mathbf{\bar p})= n E(\mathbf{\bar p}) = n \sqrt{\bar M^{2} c^{4} - \mathbf{\bar p}^{2} c^{2}}\,.\label{field:quant:disper}
 \end{equation}
 
Thus, we find that such periodic fields have the same energy spectrum as the ordinary second quantized fields after normal ordering, as shown in fig.(\ref{fig:disp:rel}.a).  
In this way we see that this quantization prescription represents a generalization  to relativistic fields of  the semi-classical quantization of a particle in a box. This approach has also interesting analogies with the Matsubara \cite{Matsubara:1955ws}, Kaluza-Klein (KK) \cite{Kaluza:1921tu,Klein:1926tv} and \emph{zitterbewegung} models.  

It must be noticed from eq.(\ref{four-period:contraint}) that the proper-time periodicity $ T_{\tau}$ is related to the mass of the field
$
T_{\tau}= T_{t} (0)={h}/{\bar M c^{2}}
 $. 
 Moreover, since the proper-time $\tau$ is proportional to the world-line parameter $s=c \tau$, we find that in this theory the world-line parameter is a cyclic variable with compactification length
 \begin{equation}
\lambda_{s}=c T_{\tau} =\frac{h}{\bar M c} \,.\label{compton:wavelength}
 \end{equation}
It actually corresponds to the Compton wavelength of matter fields. 

The proper-time periodicity (Compton time) fixes the upper bound of the time periodicity, $T_{\tau} \geq T_{t}(\mathbf{\bar p})$, since the mass is the lower bond of the energy, $\bar M c^{2} \leq \bar E(\mathbf{\bar p})$. The heavier the mass the faster the proper-time periodicity.  For instance, even a light particle such as the electron  has (in a generic reference frame) intrinsic time periodicity  faster than  $10^{-21} s$, \textit{i.e.} faster than its proper-time periodicity. It should be noted that this is many orders of magnitude away from the Cs atomic clock whose periodicity is by definition of the order of $10^{-10} s$ and it is extremely fast even if compared with  the present experimental resolution in time ($\sim 10^{-17} s$).

It is interesting to point out that since  the world-line parameter $s$ is compact, we have a deep analogy with ordinary string theory where  one of the two world-sheet parameters is supposed to be compact or periodic. Thus our field theory can be classified as a particular kind of string theory - the periodic field  eq.(\ref{field:exp:modes}) is a vibrating string. We will also see that our field theory turns out to be ``dual'' to an extra-dimensional field theory whose compactification length is $\lambda_{s}$.

 \section{Correspondence to Quantum Field Theory}
 
Now we shortly show that this description provides a remarkable matching with the canonical formulation of QM as well as with the Feynman Path Integral (FPI) formulation \cite{Dolce:2009ce,Dolce:tune,Dolce:AdSCFT,Dolce:2009cev4,Dolce:QTRF5,Dolce:2010ij,Dolce:FQXi}. 
	 The evolution along the compact time dimension is described by the so called bulk equation of motions
	 $
(\partial_{t}^{2}+\omega_{n}^{2})\phi_{n}(x,t)=0$ - for the sake of simplicity  in this section we assume a single spatial dimension $x$. 
Thus the time evolution of the energy eigenmodes can be written as first order differential equations
	  $
i\hbar\partial_{t}\phi_{n}({x},t)=E_{n}\phi_{n}({x},t)$. 
	The periodic field eq.(\ref{field:exp:modes}) is a sum of on-shell standing waves. Actually this is  the typical case where a Hilbert space can be defined.
	In fact, the energy eigenmodes form a complete set with respect to the inner product
	\begin{equation}
\left\langle \phi|\chi\right\rangle \equiv\int_{0}^{\lambda_{x}}\frac{{dx}}{{\lambda_{x}}}\phi^{*}(x)\chi(x)\,.\label{inner:prod}\end{equation}
Therefore we can define the Hilbert eigenstates as 
	 $
\left\langle {x}|\phi_{n}\right\rangle \equiv{{\phi_{n}({x})}}/{{\sqrt{\lambda_{x}}}}$. 
  On this base we can formally build a Hamiltonian operator $
\mathcal{H}\left|\phi_{n}\right\rangle\equiv\hbar\omega_{n}\left|\phi_{n}\right\rangle $ and a momentum operator $
\mathcal{P}\left|\phi_{n}\right\rangle \equiv-\hbar k_{n}\left|\phi_{n}\right\rangle$, where $k_{n}= n \bar k = n h / \lambda_{x}$.
	 Thus the time evolution of a  generic state
	 $
|\phi(0)\rangle\equiv\sum_{n}a_{n}|\phi_{n}\rangle$ 
	  is actually described by the familiar  Schr\"odinger equation  
	  \begin{equation}
i\hbar\partial_{t}|\phi(t)\rangle=\mathcal{H}|\phi(t)\rangle.\end{equation}
	 Moreover the time evolution is given by the usual time evolution operator
	$
\mathcal{U}(t';t)=\exp[{{-\frac{{i}}{\hbar}\mathcal{H}(t-t')}}] $ 
	  which  turns out to be a Markovian operator: 
	  $
\mathcal{U}(t'';t')=\prod_{m=0}^{N-1}\mathcal{U}(t'+t_{m+1};t'+t_{m}-\epsilon)$ 
where $N\epsilon=t''-t'~$.

	   From the fact that the spatial coordinate is in this theory a cyclic variable, by using the definition of the expectation value of an observable $ \hbar \partial_{x} F(x)$ between two generic initial and final states $ |\phi_{i}\rangle $ and $ |\phi_{f}\rangle$ of this Hilbert space, and after integrating by parts eq.(\ref{inner:prod}), we find
	   \begin{equation}
\left\langle \phi_{f} | \hbar \partial_{x} F(x)  |\phi_{i}\right\rangle = i \left\langle \phi_{f} | \mathcal{P} F(x) - F(x) \mathcal{P}  |\phi_{i} \right\rangle\,.
\end{equation}
	   Assuming now that the observable is such that $F(x)=x$ \cite{Feynman:1942us}  we obtain  the usual commutation relation of ordinary QM: 
	   $
[x,\mathcal{P}]=i  \hbar 
$. 
	   With this result we have checked the correspondence to canonical QM.
	 
\begin{figure}[h]
\pg{a}{}
\pg{0}{\tiny$0$}
\pg{1}{\tiny$1$}
\pg{2}{\tiny$2$}
\pg{3}{\tiny$3$}
\pg{4}{\tiny$4$}
\pg{5}{\tiny$5$}
\pg{t}{\!\!\!\!\!\small$t/T_{t}$}
\pg{x}{\small$x/ \lambda_{x}$}
\pg{p1}{ {\small$(0.6, 0.4)$}}
\pg{q}{ {\small$(4.3, 2.4)$}}
\begin{center}
\begin{minipage}{14pc}
 \includegraphics[width=14pc]{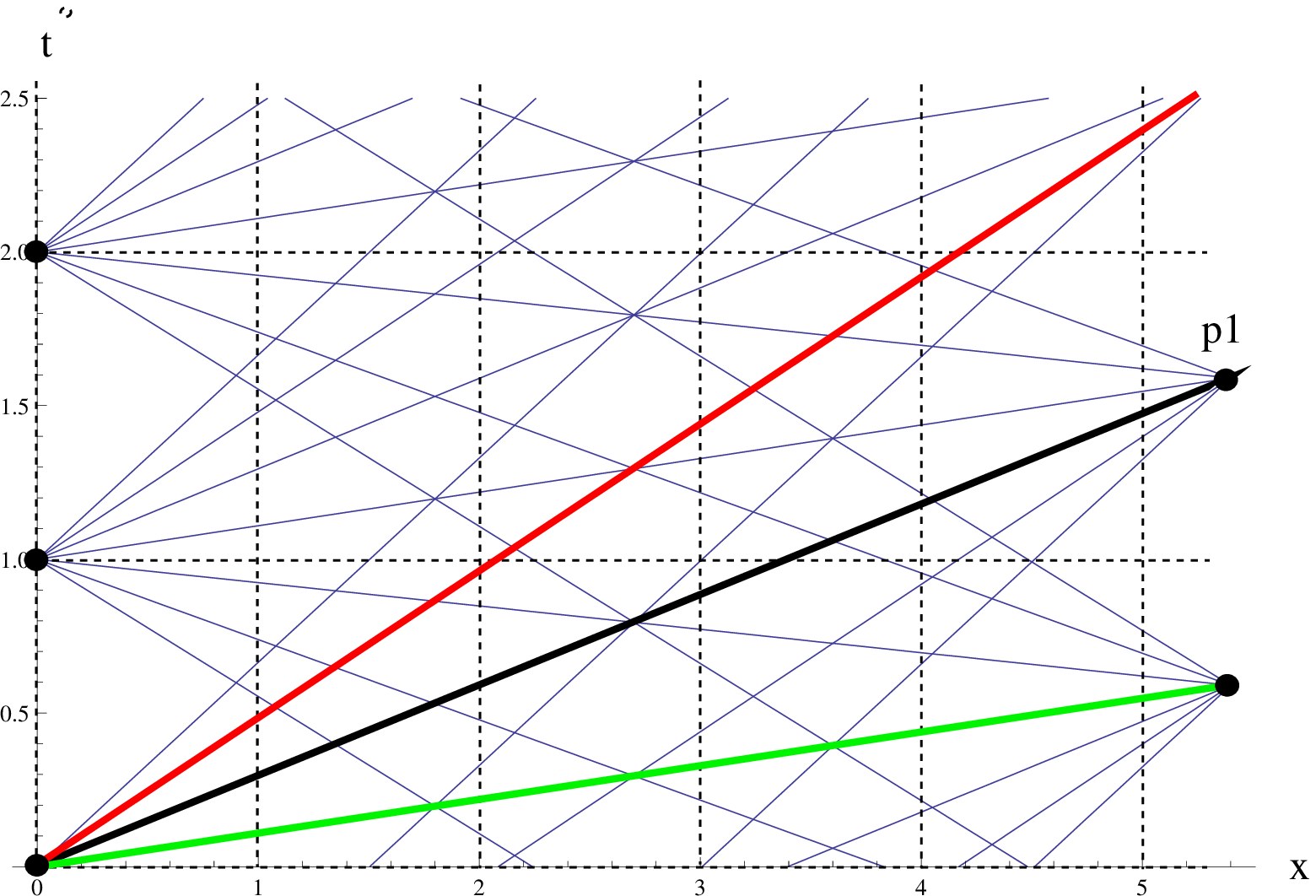}
a) 
\end{minipage}\hspace{5pc}\vspace{-1pc}%
\begin{minipage}{14pc}
 \includegraphics[width=14pc]{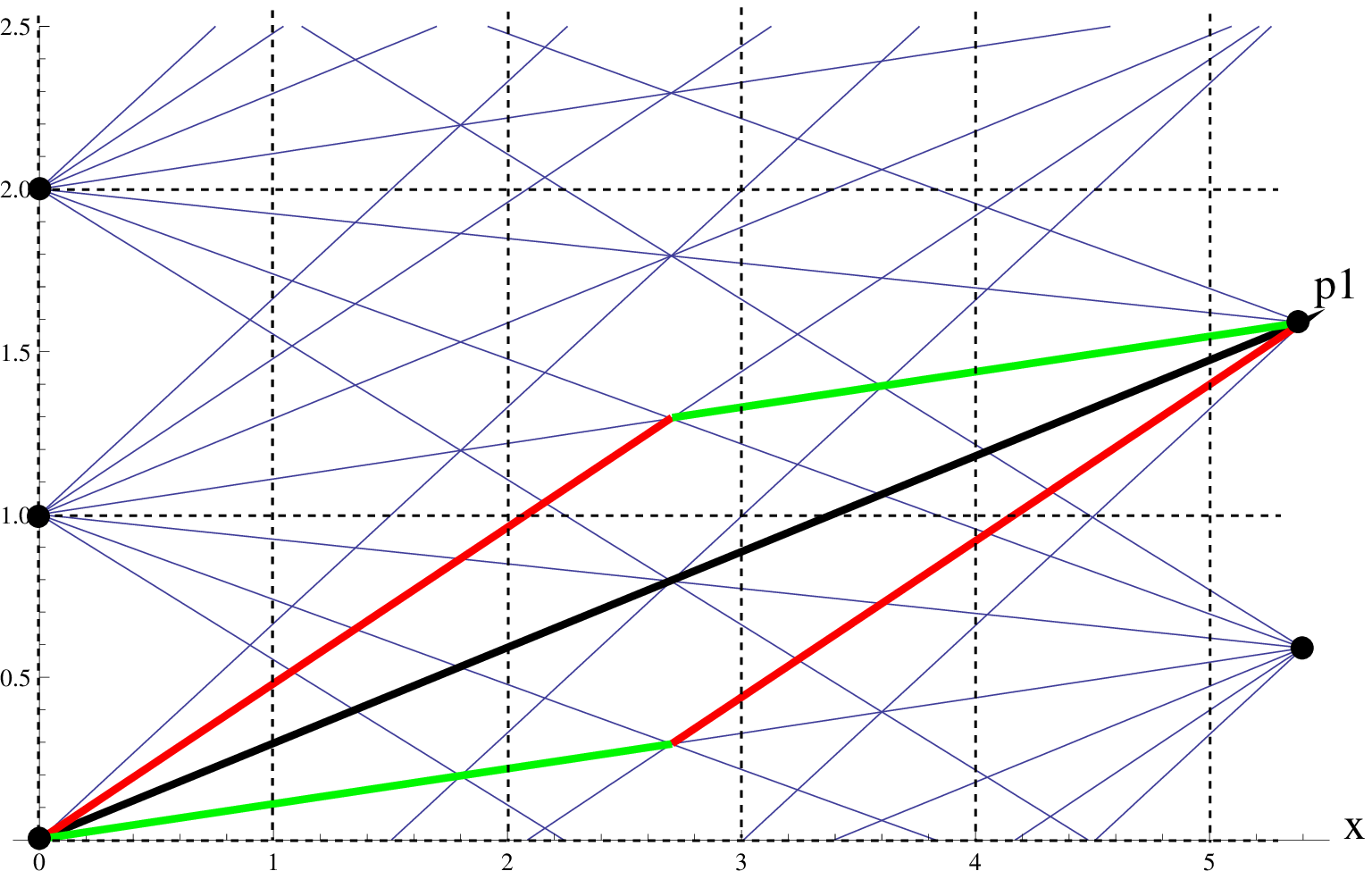}
b) 
\end{minipage} 
\end{center}
\vspace{1pc}
\begin{center}
\begin{minipage}{37pc}
\begin{center}
c) \includegraphics[width=20pc]{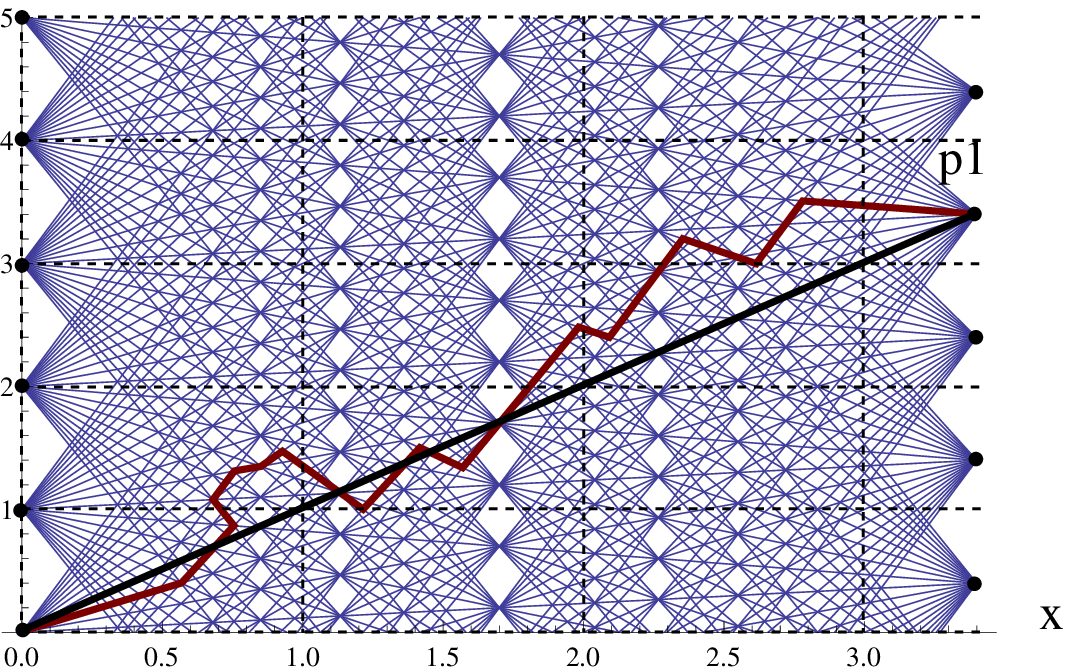}
\caption{\label{fig:period:paths}Periodic paths a) can be combined b) in order to reproduce Feynman paths c). }
\end{center}
\end{minipage}
\end{center}
\end{figure}

	Furthermore, it is possible to prove the correspondence to the FPI formulation.  In fact, it is sufficient to plug the completeness relation of the energy eigenmodes in between the elementary time evolutions of the unitary (Markovian) operator. With these elements at hand and proceeding in a completely standard way we find the usual  FPI in phase space, 
	  \begin{equation}
\mathcal{Z}=\lim_{N\rightarrow\infty}\int_{0}^{\lambda_{x}}\left(\prod_{m=1}^{N-1}dx_{m}\right)\prod_{m=0}^{N-1}\left\langle \phi\right|e^{-\frac{i}{\hbar}(\mathcal{H}\Delta\epsilon_{m}-\mathcal{P}\Delta x_{m})}\left|\phi\right\rangle\,. \label{periodic:path.integr:Oper:Fey}
\end{equation}
This important result has been obtained without any further assumption than PBCs --- see \cite{Dolce:tune} for the  generalization to  interactions. It can be  understood intuitively in a graphical way. For the sake of simplicity we again consider only time periodicity. We have to imagine a cylinder where there is an infinite set of possible classical - straight - paths with different winding numbers that  link every given initial and final points.  If we imagine opening this cylinder, the paths with different winding number are lines, but the initial and final points are no more the same --- they differ by periods --- fig.(\ref{fig:period:paths}.a). Since these evolutions are unitary and invariant under periodic translations, we can ideally cut these paths in elementary parts, translate the resulting elementary paths by periods and combine them, fig.(\ref{fig:period:paths}.b).  In the resulting combination of paths the degeneracy under periodic translations has been removed; all the resulting paths obtained from this combination of elementary periodic paths have the same initial and final points. By iterating this procedure, fig.(\ref{fig:period:paths}.c),  it is easy to see that it  reproduces --- see  \cite{Dolce:2009cev4} for a mathematical proof  --- the variation around a give classical paths of the usual Feynman formulation, see \cite{Dolce:2009ce,Dolce:2009cev4}. In this way we can figure out that with cyclic dimensions there are many possible evolutions of a field from an initial configuration to a final configuration, that can self-interfere, similarly to the non-classical path of the FPI.  However there is a fundamental conceptual difference with respect to the usual Feynman formulation, in fact now all these possible paths are classical paths, \emph{i.e.} they are equivalent to classical periodic paths. 
	 This means that in this path integral formulation it is not necessary to relax the classical variational principle in order to have self-interference. 

	 Since we have essentially standing waves, it is easy to see that there is an underlying  Heisenberg uncertainty relation \cite{Dolce:2009ce,Dolce:2009cev4}. 
	 Moreover,  since periodicity means that the only possible energy eigenmodes are those with an integer number of cycles, we obtain the  Bohr-Sommerfeld quantization condition (for instance the periodicity condition $E_{n} T_{t} = n h $ can be more in general written as $\oint E_{n} d t = n h$). This allows us to solve many non-relativistic quantum problems \cite{Dolce:2009ce,Dolce:2009cev4}.  
	
	The non-quantum limit of a massive field, that is the  non-relativistic single particle description, is obtained by setting the mass to infinity so that, as shown in \cite{Dolce:2009ce,Dolce:2009cev4}, in an effective classical limit, only the first level of the energy spectrum must be considered. This leads to a consistent interpretation of the wave-particle duality and of the double slit experiment. The quantities describing the first energy level are addressed by the bar sign. For instance, the Lagrangian of the fundamental mode $\bar \Phi(x)$ is $\mathcal{\bar L}_{\lambda_{s}}(\partial_{\mu}\bar \Phi(x),\bar \Phi(x))$.

On the other hand a massless field has infinite Compton wavelength and thus an infinite proper-time periodicity, fig.(\ref{fig:disp:rel}.d). Its  quantum limit  is at high frequency where the PBCs are important. In this limit we have discretized energy spectrum, in agreement with the ordinary description  of the black-body radiation (no UV catastrophe), fig.(\ref{fig:disp:rel}.c). The opposite limit is when the time periodicity tends to infinity and we get back a continuous energy spectrum.

A remarkable property of this theory is that  QM emerges from PBCs without involving any local-hidden-variable. This suggests that the theory can in principle violates  Bell's inequality and  we can actually speak about determinism. 
The idea is similar to the stroboscopic quantization or to the 't Hooft determinism. In fact, 
the intrinsic time periodicities are typically  extremely fast. Thus, we inevitably have a too low  revolution in time, so that at every observation the system appears in an aleatoric phase of its evolution, just like a  clock  observed under a stroboscopic light.  
Thus, the de Broglie intrinsic clocks of elementary particles can be imagined as a ``de Broglie deterministic dice'', that is dice rolling with time periodicities $T_{t}$. Since such dice roll too fast with respect to our time resolution, we can only predict the outcomes statistically.  The results presented in this section show us that the statistical description associated to intrinsically fast periodic phenomena actually matches ordinary QM.

 \section{Matching with relativistic geometrodynamics}

To introduce interactions  we must bear in mind that the four-periodicity $T^{\mu}$ is fixed by the inverse of the four-momentum $\bar p_{\mu}$ according to the de Broglie-Planck relation eq.(\ref{deBroglie:Planck:rel}). 

As already said, an isolated  elementary system  (\textit{i.e.} free  field) has persistent four-momentum.  On the other hand, an elementary system under a generic interaction scheme  can be described locally in terms of corresponding  variations  of its four-momentum  with respect  to the free case 
\begin{equation}
\bar{p}_{\mu}\rightarrow\bar{p}'_{\mu}(x)=e_{\mu}^{a}(x)\bar{p}_{a}\,.\label{eq:deform:4mom:generic:int}
\end{equation}
In other words we describe interactions in terms of the  so call tetrad (or virebein)  $e_{\mu}^{a}(x)$. Thus the interaction scheme eq.(\ref{eq:deform:4mom:generic:int}) turns out to be  encoded in the corresponding local variation of the space-time periodicities
\begin{equation}
T^{\mu}\rightarrow T'^{\mu}(x)=e_{a}^{\mu}(x)T^{a}\,.\label{eq:deform:4period:generic:int}
\end{equation}
This is the corresponding deformation of the compactification lengths of a periodic field as well (in the case of sufficiently smooths or exponential deformations we approximate the compactification length with the local periodicity). 
Roughly speaking, interactions can be thought of as stretching of the compact dimensions of the theory. Equivalently, the interaction eq.(\ref{eq:deform:4mom:generic:int}) turns out to be encoded in the corresponding curved space-time background 
\begin{equation}
\eta_{\mu\nu}\rightarrow g_{\mu\nu}(x)=e_{\mu}^{a}(x)e_{\nu}^{b}(x)\eta_{ab}\,.\label{eq:deform:metric:generic:int}
\end{equation}

This result  can be double checked by considering the transformation of space-time variables 
\begin{equation}
dx_{\mu}\rightarrow dx'_{\mu}(x)=e_{\mu}^{a}(x)dx_{a}\,.\label{eq:deform:mesure:generic:int}
\end{equation} 
 This transformation of variables (diffeomorphism) with determinant of the Jacobian $\sqrt{-g(x)}$,  the free action eq.(\ref{free:act}) turns out to be 
\begin{equation}
\mathcal{S}_{\lambda_{s}}\sim\oint^{e_{a}^{\mu}(x)T^{a}}d^{4}x\sqrt{-g(x)}\mathcal{L}_{\lambda_{s}}(e_{\mu}^{a}(x)\partial_{a}\Phi(x),\Phi(x))\,.\label{eq:defom:action:generic:int}\end{equation}
Therefore, the periodic field which minimizes this action has four-periodicity $T'^{\mu}$, eq.(\ref{eq:deform:4period:generic:int}), or equivalently has four-momentum $\bar p_{\mu}$, eq.(\ref{eq:deform:4mom:generic:int}).   
 We conclude that a field under the interaction scheme eq.(\ref{eq:deform:4mom:generic:int}) is described locally by the  solutions of the bulk equations of motion with PBCs  on the deformed compact background eq.(\ref{eq:deform:metric:generic:int}) and compactification lengths eq.(\ref{eq:deform:4period:generic:int}).

This geometrodynamical approach to interactions is interesting because it actually mimics the usual geometrodynamical approach of GR. In fact, if we suppose a weak Newton potential
$ 
V(\mathbf{x})=-{GM_{\odot}}/{|\mathbf{x}| c^{2}}\ll1\label{eq:low:grav:pot}
$, 
 we find that the energy in a gravitational well  varies (with respect to the free case) as
$ 
\bar{E}\rightarrow\bar{E}'\sim\left(1+{{GM_{\odot}}}/{|\mathbf{x}| c^{2}}\right)\bar{E}
$. 
According to eq.(\ref{eq:deform:4period:generic:int}) or eq(\ref{deBroglie:Planck:rel}), this means that the de Broglie clocks in a gravitational well are slower with respect to the free clocks 
$
T_{t}\rightarrow T_{t}'\sim\left(1-{{GM_{\odot}}}/{|\mathbf{x}|c^{2}}\right)T_{t}
$. 
 Thus we have a gravitational redshift
 $
\bar{\omega}\rightarrow\bar{\omega}'\sim\left(1+{{GM_{\odot}}}/{|\mathbf{x}|c^{2}}\right)\bar{\omega}
$. 
With this schematization of interactions we have retrieved two important predictions of GR. 

Besides this, \cite{Ohanian:1995uu}, we must also consider  the analogous variation of spatial momentum 
$ 
\mathbf{\bar{p}}\rightarrow\mathbf{\bar{p}}'\sim\left(1+{{GM_{\odot}}}/{|\mathbf{x}| c^{2}}\right)\mathbf{\bar{p}}
$, 
and the corresponding variation of spatial periodicities  
$ 
\vec{\mathbf{\lambda}}\rightarrow \vec{\mathbf{\lambda}}'\sim\left(1-{{GM_{\odot}}}/{|\mathbf{x}|c^{2}}\right)\vec{\mathbf{\lambda}}
$. 
According to the relation eq.(\ref{eq:deform:metric:generic:int}), this means that the weak Newtonian interaction turns out to be encoded in the usual Schwarzschild metric
\begin{equation}
ds^{2}=\left(1-\frac{{2 GM_{\odot}}}{|\mathbf{x}| c^{2}}\right)dt^{2}-\left(1+\frac{{2 GM_{\odot}}}{|\mathbf{x}| c^{2}}\right)d|\mathbf{x}|^{2} - |\mathbf{x}|^{2} d \Omega^{2}\,,
\end{equation}
We have found that the  geometrodynamical approach to interactions actually can be used to describe linearized gravity and that the geometrodynamics of the compact space-time dimensions correspond to the usual relativistic ones. 

As well known, see for instance \cite{Ohanian:1995uu}, it is possible to  retrieve  ordinary GR from a linear formulation  by including self-interactions. More naively, as we will show in detail in a forthcoming paper, 
 we can add ``by hand'' a kinetic term to the Lagrangian (with appropriate coupling), in order to describe the dynamics of the metric $g_{\mu\nu}$ which is the new \textit{d.o.f.} of the theory eq.(\ref{eq:defom:action:generic:int}). Thus, neglecting quantum corrections, we can replace the classical limit  $\sqrt{-g} \mathcal{\bar L}_{\lambda_{s}}$ of the Lagrangian in eq.(\ref{eq:defom:action:generic:int}) with the Hilbert-Einstein Lagrangian
\begin{equation}
{ \mathcal{\bar L}}_{HE}=\sqrt{-g} \left[ -\frac{g^{\mu\nu} \mathcal{R}_{\mu\nu}}{16 \pi G_{N} } + \mathcal{\bar L}_{\lambda_{s}}(e_{\mu}^{a}\partial_{a}\bar \Phi(x),\bar \Phi(x))\right ] \,.\label{eq:defom:action:generic:int}\end{equation}
It can be shown that this procedure is parallel to what we usually do in electromagnetism when we add the term $- F_{\mu\nu} F^{\mu\nu}/4 e^{2}$  to describe the kinematics of the gauge field. Intuitively, because of  its geometrical meaning, the Ricci tensor is the correct  mathematical object to describe the variations of the space-time compactification lengths in different space-time points. Bearing in mind eq.(\ref{deBroglie:Planck:rel}),  we note that actually such a kinetic term must encode the content of four-momentum in different space-time points.  By varying the metric (neglecting issues related to the variation of boundary terms of the Hilbert-Einstein action and related BCs \cite{springerlink:10.1007/BF01889475})  eq.(\ref{eq:defom:action:generic:int}) yields  the usual Einstein equation
$ 
\mathcal{R}^{\mu\nu}=-8 \pi G_{N} \mathcal{T}^{\mu\nu}$. 
With these simple and heuristic arguments we have shown that  field theory in compact space-time is in agreement not only with special relativity but also with GR. 
In forthcoming papers 
we will show that, by writing eq.(\ref{eq:deform:4mom:generic:int}) as a minimal substitution, such a geometrodynamical approach to interactions can be also used to describe ordinary gauge interactions. Gauge fields will turn out to ``tune'' the variation of periodicities,  allowing a semi-classical interpretation of superconductivity \cite{
Dolce:tune}.

\section{``Virtual'' Extra Dimension (VXD)}

Our geometrodynamical description of interactions provides an intuitive interpretation of AdS/CFT whose essential meaning is summarized by the Witten's words: ``In this description, quantum phenomena [...] are coded in classical geometry'' \cite{Witten:1998zw}. 
To understand this point  we should note that our field theory in compact space-time  is ``dual'' to an extra dimensional field theory, see \cite{Dolce:2009ce,Dolce:2009cev4,Dolce:AdSCFT}. For instance we may note that  the energy spectrum eq.(\ref{field:quant:disper})  in the rest frame ($\mathbf{\bar p}\equiv 0$) reproduces the usual KK mass tower $M_{n} = E_{n}(0)/ c^{2} = n \bar M = n h / \lambda_{s} c$, see fig.(1.a).  Practically, the recipe to obtain  our periodic fields  from a corresponding  extra dimensional field on a flat five dimensional  metric ($d S^{2} = d x_{\mu} d x^{\mu} - d s^{2}$) and with zero five dimensional mass ($d S^{2} = 0$), is to  identify the compact extra dimension $s$  with our compact world-line parameter $s= c \tau$ and integrating it out, see \cite{Dolce:2009ce,Dolce:2009cev4,Dolce:AdSCFT} for more detail.  In this way we get back the usual four-dimensional Minkowskian metric $d s^{2}= d x_{\mu} d x^{\mu}$. In particular we find that the compactification length of the eXtra Dimension (XD) corresponds to the compactification length $\lambda_{s}$ of the compact world-line parameter $s$, \textit{i.e.} to the Compton wavelength eq.(\ref{compton:wavelength}). 
When we identify the XD with the world-line parameter we say that it is a ``Virtual'' eXtra Dimension (VXD) and the KK mode are the ``virtual'' modes of the field\footnote{Note that in the original Kaluza's formulation \cite{Kaluza:1921tu} the XD was introduced not like an ``real'' XD but as a ``mathematical trick'', whereas in the original Klein's proposal \cite{Klein:1926tv} the assumption of cyclic (\textit{i.e.} compact with PBCs) XD was used as a semi-classical quantization condition.}.
Thus, in our theory, the KK modes  are not different elementary fields as in the usual KK scenario but collective modes of the same 4D periodic field, similarly to the holographic prescription.  They play the role of energy eigenmodes.  
Finally, since periodic fields reproduce QM and, on the other hand, they turn out to be dual to  extra dimensional fields, we infer that the classical geometrodynamics on  a VXD could be used to describe quantum behaviors, similarly to AdS/CFT.

Shortly, in the approximation of massless fields, to describe the interaction scheme eq.(\ref{eq:deform:4mom:generic:int}) through the VXD formalism  we have to generalize the deformed metric eq.(\ref{eq:deform:metric:generic:int}) to the following deformed metric in VXD 
\begin{equation}
g_{M N}(s)\sim\left(\begin{array}{cc}
g_{\mu\nu}(x(s)) & 0\\
0 & 1\end{array}\right)\,,\label{VXD:deform:metric}
\end{equation}
where the capital letters label the 5D Lorentz indices.
 Thus we expect to find out that the classical geometrodynamics of the field in such a deformed metric with VXD reproduces the quantum behaviors of the corresponding interaction scheme eq.(\ref{eq:deform:4mom:generic:int}). Similarly to the Maldacena conjecture, this correspondence can be summarized by the mnemonic relation 
 \begin{equation}
\mathcal{Z}=\int_{0}^{\lambda_{x}}\!\mathcal{D}x ~e^{\frac{i}{\hbar}\mathcal{S}(s,s')[\dot{x},x]}\leftrightarrow~e^{\frac{i}{\hbar}\mathcal{S}^{VXD}(s,s')}\,.\label{VXD:QFT:corr}\end{equation}

For instance \cite{Dolce:2009cev4},  we can consider a collider experiment where the Quark-Gluon-Plasma (QGP) can be  approximated  as a volume of massless fields with high four-momentum\footnote{This example is of particular  interest because of the recent data from LHC}. 
As predicted by a simple Bjorken hydrodynamical model \cite{Magas:2003yp}, or by thinking of the analogy of QCD with a thermodynamical system \cite{Satz:2008kb}, the four-momentum of the fields decays exponentially and conformally as the QGP radiates hadronically and electromagnetically\footnote{The  simplest Bjorken kinetic model  \cite{Magas:2003yp} describing the exponential gradient of temperature of the QGP freeze-out can be regarded  as the analogous of  Newton's law of cooling in QCD thermodynamics.}:
$ 
\bar{p}_{\mu}\rightarrow\bar{p}'_{\mu}(s)\simeq e^{-ks}\bar{p}_{\mu}\,.\label{eq:deform:4mom:QGP}
$  - in natural units. 
 The QGP freeze-out is, in first approximation, encoded in the tetrad $e^{a}_{\mu}(s)\simeq \delta^{a}_{\mu} e^{-ks}$. 
This means that  the space-time periodicities of the fields have a conformal and exponential dilatation
$ 
T^{\mu}\rightarrow T'^{\mu}(s)\simeq e^{ks}T{}^{\mu}\label{eq:deform:4period:QGP}
$. Indeed the time periodicity $T'_{t}(s)$ is the usual conformal parameter of AdS/CFT  
and  the QGP freeze-out is effectively be described by the virtual AdS metric
\begin{equation}
dS^{2}\simeq e^{-2ks}dx{}_{\mu}dx{}^{\mu}-ds^{2}\equiv0~.
\end{equation}
According to our description,  it is easy to prove \cite{Pomarol:2000hp,ArkaniHamed:2000ds} that the propagation of  classical  fields an warped virtual metric with infinite VXD describes the quantum behavior of a massless theory such as CFT. 
A finite VXD gives mass to the virtual KK modes breaking the conformal invariance. Their collective classical propagation can described by using holography \cite{Pomarol:2000hp,ArkaniHamed:2000ds}, \footnote{To evaluate the low energies effective propagator we assume Neumann BCs at the UV scales $\Lambda$ and boundary field $A_{\mu}(q)$ at the IR scale $\mu$, in the hypothesis of Euclidean momentum $q$ such that $\Lambda \gg  |q| \gg \mu$. 
	 \cite{ArkaniHamed:2000ds,Dolce:2009cev4}.} 	   
\begin{equation}
\Pi^{Holo}(q^{2})\sim-\frac{q^{2}}{2kg_{5}^{2}}\log\frac{q^{2}}{\Lambda^{2}}~,\end{equation}
which assuming $1/k g_{5^{2}} = N_{c } / 12 \pi^{2}$, can be matched to the quantum two point function of QCD. The mass spectrum of the virtual KK modes can be interpreted as an hadronic spectrum. Thus we  obtain semi-classically the quantum running of the strong coupling constant
\begin{equation}
\frac{1}{e_{eff}^{2}(q)}\sim\frac{1}{e^{2}}-\frac{{N_{c}}}{{12\pi^{2}}}\log\frac{q}{\Lambda}~.\end{equation}
These are  fundamental characteristics of the  AdS/QCD models.

\section{Conclusions}

To understand the conceptual meaning of this theory we have to note that time can only be defined
by counting the number of cycles of phenomena \textit{supposed} to be periodic.
Only by assuming periodicity we can ensure that the duration of the unit
of time is always the same; in the past, in the present and in the future, \cite{Einstein:1910}.
In particular, the usual - non compact - time axis ($t\in \mathbb{R}$) is defined with reference to the
Cs-133 clock whose characteristic periodicity 
 $\sim 10^{-10}s$. Hence, for a consistent formulation of
 natural laws we believe  that there must be an assumption of periodicity in physics. We explicitly introduce this assumption by imposing
on every free field its de Broglie intrinsic periodicity as a constraint.
As shown in detail in \cite{Dolce:2009ce}, see also \cite{Dolce:2009cev4,Dolce:QTRF5,Dolce:2010ij}, and as we have summarized
in this paper, under this assumption ordinary QM
emerges as  an effective description of the underlying deterministic field theory by
a process of ``information loss''. It is important
to note that in our case there are no local-hidden-variables, time being
  a physical variable that can not be eliminated from the theory, and 
the PBCs being an element of non locality. 
This suggests that the theory is not constrained by Bell's or similar
theorems; it can in principle violate Bell's inequality and reproduce the predictions of
QM. This allows us to speculate about a scenario where, by observing the de Broglie dice of electrons  with  time resolution  greater that the ZHz, it could be possible to resolve the underlying deterministic cyclic electrodynamics and in principle predict the outcomes of the quantum dice. 

These cyclic fields can therefore be  identified with the so call  ``de Broglie
internal clocks'' or ``de Broglie periodic phenomena''. Similarly to a calendar or to a stopwatch where every moment in time is determined by the combination of the phases of periodic cycles (typically: years, months, days, hours, minutes and seconds), every
value of  our external temporal axes  (defined in terms of the ``ticks'' of the Cs-133 atomic clock) is characterized by a unique combination
of the phases of the de Broglie internal clocks,  \textit{i.e.} by the  ``ticks''  of all the
elementary particles constituting the system under investigation.   In this scenario the long time scales are provided by massless fields with low frequencies (low energies implies long time periodicities). 
The combination of two or more clocks - \emph{i.e.}  a non elementary system -
with irrational ratio of periodicities, gives ergodic - or even more chaotic - evolutions. 
Furthermore  the clocks can
vary their periodicity through interactions (exchange of energy) as their periodicities
depend on the reference systems according to the relativistic laws. 
Once  the de Broglie internal clocks are fixed to be clockwise or anticlockwise, the flow of time is determined by the combinations of their ``ticks'' and the variations of their periodicities through interactions. This
means that the external time axis can be in principle dropped and the flow of time
can be effectively described in terms of the ``ticks'' of these
de Broglie internal clocks.
In this scenario the flow of time doesn't
depend on the assumption of clockwise or anticlockwise rotations of
the de Broglie internal clocks. The  resulting  formulation of the flow of time is particularly interesting for  the problem of the time arrow in physics.
\vspace{-1em}
\section*{References}
\bibliographystyle{iopart-num}
\providecommand{\newblock}{}

\end{document}